\documentstyle[epsf,prl,twocolumn,aps]{revtex}
\begin{document}
\draft
\preprint{HEP/123-qed}



\twocolumn[
\hsize\textwidth\columnwidth\hsize\csname
@twocolumnfalse\endcsname
\title{Interface superconductivity in the eutectic Sr$_2$RuO$_4$-Ru: 
3-K phase of Sr$_2$RuO$_4$}

\author{Hiroshi Yaguchi}
\address{Department of Physics, Graduate School of Science, Kyoto 
University, Kyoto 606-8502, Japan \\
CREST, Japan Science and Technology Corporation, Kawaguchi, 
Saitama 332-0012, Japan \\
}

\author{Masahiko Wada and Takashi Akima}
\address{Department of Physics, Graduate School of Science, Kyoto 
University, Kyoto 606-8502, Japan \\
}

\author{Yoshiteru Maeno}
\address{Department of Physics, Graduate School of Science, Kyoto 
University, Kyoto 606-8502, Japan \\
CREST, Japan Science and Technology Corporation, Kawaguchi, 
Saitama 332-0012, Japan \\
Kyoto University International Innovation Center, Kyoto 
606-8501, Japan}

\author{Takehiko Ishiguro}
\address{Department of Physics, Graduate School of Science, Kyoto 
University, Kyoto 606-8502, Japan \\
CREST, Japan Science and Technology Corporation, Kawaguchi, 
Saitama 332-0012, Japan \\
}

\date{\today}

\maketitle

\begin{abstract}
\quad The eutectic system Sr$_2$RuO$_4$-Ru is referred to as the 3-K 
phase of the spin-triplet supeconductor Sr$_2$RuO$_4$ because of its 
enhanced superconducting transition temperature 
$T_{\rm c}$ of $\sim$3 K. 
We have investigated the field-temperature ($H$-$T$) phase diagram 
of the 3-K phase for fields parallel and perpendicular to the 
ab-plane of Sr$_2$RuO$_4$, using out-of-plane resistivity measurements. 
We have found an upturn curvature in the $H_{\rm c2}(T)$ curve for $H {\parallel}$ c, and a rather gradual temperature dependence of $H_{\rm c2}$ close to $T_{\rm c}$ 
for both $H {\parallel}$ ab and $H {\parallel}$ c. 
We have also investigated the dependence of $H_{\rm c2}$ on the angle between the field and the ab-plane at several temperatures. Fitting the Ginzburg-Landau effective-mass model apparently fails to reproduce the angle dependence, particularly near $H {\parallel}$ c and at low temperatures. 
We propose that all of these charecteric features can be explained, at least in a qualitative fashion, on the basis of a theory by Sigrist and Monien that assumes 
surface superconductivity with a two-component order parameter 
occurring at the interface between Sr$_{2}$RuO$_{4}$ and Ru 
inclusions. This provides evidence of the chiral state postulated for the 1.5-K phase by several experiments.

\end{abstract}
\pacs{PACS numbers: 74.25.Dw, 74.70.Pq, 74.62.Bf, 74.80.-g}


]

\narrowtext
 
\section{Introduction}
\quad Sr$_2$RuO$_4$ is the first layered perovskite superconductor without copper  \cite{discovery}; it is isostructual to the cuprate 
high-temperature superconductor La$_{2-x}$Ba$_x$CuO$_4$.
This is one of the reasons that Sr$_2$RuO$_4$ 
has attracted great attention \cite{review} despite its superconducting transition 
temperature $T_{\rm c}$ being rather low (ideally 1.5 K) \cite{impurity,defect}.
In fact, a number of theoretical and experimental studies 
subsequently made have revealed its unconventional natures. 
More importantly, it is now well established 
that Sr$_2$RuO$_4$ is a spin-triplet superconductor, 
in contrast to the spin-singlet {\it d}-wave pairing in high $T_{\rm c}$ cuprates \cite{review}. This has been first confirmed by $^{17}$O-NMR measurements \cite{NMR}. 
The Knight shift is unaffected by the superconducting transition, 
strongly suggestive of spin-triplet pairing with the spin of Cooper pairs lying within the ab-plane \cite{NMR}.
Also, the observation of spontaneous magnetic moments 
accompanying the superconducting state indicates broken time-reversal 
symmetry \cite{muSR}, suggesting a two-component order parameter with a relative phase of ${\pi}/2$.
Subsequent experiments of Ru-NMR \cite{RuNMR} and polarised neutron scattering \cite{polarised}, both of which measure the Knight shift, support the $^{17}$O-NMR measurements \cite{NMR}. A detailed small-angle neutron scattering study has revealed the vortex field distribution, which cannot be understood without the two-component order parameter \cite{SANS}.
These results constrain the basic form of the vector order parameter to be {\bf{\textit{d}}}(\textbf{\textit{k}}) = {\bf{\textit{z}}}$\Delta_{0}(k_x + ik_y)$, called a chiral state. 

\quad Since the Fermi surface consists of three cylindrical (quasi-two dimensional) sheets~\cite{dHvA,AMRO}, the above vector order parameter leads to an isotropic gap ($\Delta (\textbf{\textit{k}}) = \Delta_{0}\sqrt{k_x^2 + k_y^2}$~). However, a number of experimental results~\cite{Cp,NQR,penetration,tanatar,izawa,suzuki,ultra} have revealed the power-law temperature dependece of various thermodynamic quantities and thus strongly suggest lines of nodes in the superconducting gap. 
This fact postulates modifications to be made to the basic form {\bf{\textit{d}}}(\textbf{\textit{k}}) = {\bf{\textit{z}}}$\Delta_{0}(k_x + ik_y)$. In fact, several theories have been proposed to reconcile the descripancy between those experimental facts and the vector order parameter {\bf{\textit{d}}}(\textbf{\textit{k}}) = {\bf{\textit{z}}}$\Delta_{0}(k_x + ik_y)$ ~\cite {zhitomirsky,nomura}. These theories take into account the orbital dependent superconductivity~\cite{ODS} and propose very strong in-plane anisotropy ~\cite{nomura} 
or horizontal lines of nodes~\cite {zhitomirsky} 
in the superconducting gap to explain the power-law temperature dependence in thermodynamic data.

\quad Amongst several remarkable features related to Sr$_{2}$RuO$_{4}$, an enhancement of $T_{\rm c}$ in the Sr$_{2}$RuO$_{4}$-Ru 
eutectic system, is rather surprising. 
This enhancement was found during the course of the optimisation of sample growth.
Whilst the ideal $T_{\rm c}$ of Sr$_{2}$RuO$_{4}$ turned out to be 1.5 K\cite{impurity,defect}, the a.c. susceptibility of certain batches was known to exhibit rather weak diamagnetism at a considerably higher onset temperature of about 3 K. A clear resistance drop, below which the resistance does not necessarily fall to zero, was also observed at a very close temperature. 
As discussed in ref.~\cite{maeno3K}, Maeno {\it et al.} established that these observations are indeed due to superconductivity and, as a result of careful investigations into the material origin, that it reproducibly occurs in the Sr$_{2}$RuO$_{4}$-Ru eutectic.

\quad The eutectic system, a two-phase composite structure of a single-crystalline Sr$_{2}$RuO$_{4}$ matrix and lamellar microdomains of ruthenium metal embedded in it~\cite{maeno3K}, is obtained by the same method as Sr$_{2}$RuO$_{4}$ but with excess of Ru and/or at a faster growth speed~\cite{growth}. 
The top panel of Fig. 1 shows an optical microscopy picture of a polished surface parallel to the RuO$_{2}$ plane. Typical dimensions of lamellae are 1 $\mu$m in thickness and 1-30 $\mu$m in length and width; the separation between adjacent lamellae is of the order of 10 $\mu$m.
Although the appearance of the Ru-inclusions may depend on growth conditions, the density of Ru-inclusions should be uniquely determined by the composition at the eutectic point of Sr$_{2}$RuO$_{4}$ and Ru. In the top panel of Fig. 1, lamellae apparently line up along a certain direction, but empirically, there is no particular preferred orientation relative to the crystallographical axes. The diretion often varies even within a small piece of single crystal of Sr$_{2}$RuO$_{4}$.

\quad Such a eutectic system shows a broad superconducting transition with an onset of about 3 K. On further cooling, this transition is followed by the original superconducting transition of Sr$_{2}$RuO$_{4}$ at 1.5 K. The higher $T_{\rm c}$ superconductivity is called the 3-K phase and the original lower $T_{\rm c}$ superconductivity is referred to as the 1.5-K phase. The manifestations of 3-K phase superconductivity in resistance and a.c. suceptibility suggest that the superconductivity is inhomogeneous and filamentary. 
In addition, 3-K phase superconductivity is considered to be essentially sustained in Sr$_{2}$RuO$_{4}$.
Because the upper critical field of the 3-K phase is the highest (lowest) 
when the applied field is parallel (perpendicular) to the ab-plane of Sr$_{2}$RuO$_{4}$ ~\cite{maeno3K}. ({\it i.e.} The anisotropy of the upper critical field reflects the crystallographical directions of Sr$_{2}$RuO$_{4}$.)

\quad In addition to the enhancement of $T_{\rm c}$ to $\sim$3 K, the field-temperature phase diagram of the 3-K phase has intriguing properties. Earlier work of resistive measurements ~\cite{ando3K} has revealed clear hysteresis of the upper critical field $H_{\rm c2}$ in magnetic fields parallel to the ab-plane at low temperatures: 
Two distinctly different $H_{\rm c2}$'s are obtained when the applied magnetic field (or the temperature) is swept upwards and downwards.
Also the $H_{\rm c2} (T)$ curve for $H {\parallel}$ c looks rather unusual, being nearly a straight line or possibly concave upwards ~\cite{maeno3K,ando3K}.

\quad Neither theoretically nor experimentally has very much been known about the 3-K phase thus far. However, recent tunnelling measurements on c-axis junctions of the Sr$_{2}$RuO$_{4}$-Ru eutectic have observed zero bias conductance peaks ~\cite{tunnelling}, which is a hallmark of unconventional superconductivity~\cite{tanaka}. Therefore, the supeconductivity in the 3-K phase is also unconventional and probably originates from the triplet pairing of Sr$_{2}$RuO$_{4}$. Also Sigrist and Monien (SM) \cite{theory} have recently proposed a phenomelogical theory that assumes surface spin-triplet superconductivity at the Sr$_{2}$RuO$_{4}$-Ru interface although the theory does not consider the mechanism of the enhanced superconductivity.

\quad In the present work, we have studied the field-temperature phase diagram to higher precision than the previous work in ref. \cite{ando3K} for a further discussion. We have also investigated the dependence of the upper critical field on the angle between the applied field and the ab-plane. We will discuss these results with the help of SM's theory \cite{theory}. Besides, we have measured the specific heat to obtain thermodynamic evidence for non-bulk superconductivity, which supports an assumption of their theory.

\section{Experiment}
\quad The eutectic samples of Sr$_2$RuO$_4$-Ru used in this study 
were grown by a floating zone method. The Ru inclusions were lamellate with typical dimensions of 1 $\mu$m ${\times}$10 $\mu$m ${\times}$30 $\mu$m.  A surface parallel to the ab-plane of the sample used for resistive measurements is shown in the top panel of Fig. 1. There were no particular directions along which lamellae preferrably line up throughout the whole sample. Details of the crystal growth are described in ref. \cite{maeno3K}. We have measured the resistivity as a function of magnetic field or temperature to determine $H_{\rm c2}(T)$ and $T_{\rm c} (H)$. 
The dimensions of the sample, cut from a crystalline rod, 
were 0.96 ${\times}$1.04 mm$^2$ in the ab-plane 
and 0.58 mm along the c-axis. 
We employed a lock-in technique at 137 Hz with a current of 0.5 mA along the c-axis. Low temperatures down to 60 mK were reached by means of a $^{3}$He cryostat or a dilution refrigerator. Magnetic fields of up to 5 T were generated by a superconducting solenoid.
The sample was mounted in a rotator that enabled the angle between the ab-plane and the applied magnetic field to be changed.
\quad We have also measured the specific heat by a relaxation method (Quantum Design, model PPMS) down to 0.4 K. The sample for the specific heat measurement was cut from the same crystalline rod as used for the resistive measurements, and weighed at about 11 mg.

\section{Results and Discussion}

\subsection{$H$-$T$ Phase Diagram}
\quad Figure 2 shows typical traces of the resistance as a function of magnetic field or temperature. The transition point has been defined as the inflection point associated with the superconducting transition to the 3-K phase. Figure 2 also demonstrates that the transition points determined from $H_{\rm c2}(T)$ and $T_{\rm c} (H)$ show a good agreement.

\quad Figure 3 shows the resultant field-temperature ($H$-$T$) phase diagram of the 3-K phase of Sr$_{2}$RuO$_{4}$ for fields parallel to the ab-plane and the c-axis. This phase diagram contains a considerably larger number of transition points than those in ref. \cite{ando3K}, which makes possible a more detailed discussion. We, however, note that both phase diagrams appear to be very similar and consistent. 

\quad As seen in the phase diagram of ref. \cite{ando3K}, there are two branches, corresponding to up- and down-sweeps, below $\sim$1.2 K for $H {\parallel}$ ab. This is a consequence of the hysteresis of $H_{\rm c2}$ mentioned in the introduction. 
As discussed in ref. \cite{ando3K}, two possiblities may be envisaged for the hysteresis. One is that the magnetic field effectively applied to the region responsible for the 3-K phase superconductivity is hysteretic. The second one is that the hysteresis of $H_{\rm c2}$ is instrinsic ({\it i.e.} due to a first order transition). Obviously, the latter case is even more interesting as the superconducting transition (type II) in magnetic fields is normally second order \cite{CeCoIn5}. Possible interpretations for the latter case will include a first order transition due to spin depairing \cite{clogston}. However, we point out that this is  irrelevant to the spin-triplet state we suggest (chiral state). 

\quad The present study has revealed that the lower branch (down sweep of field or temperature) of the $H_{\rm c2}(T)$ curve for $H {\parallel}$ ab is nearly flat, which seems to be rather unusual. This finding may contribute to a further understanding of the hysteretic behaviour of $H_{\rm c2}$. In addition to the hysteresis, we note two prominent features confirmed only in the $H$-$T$ phase diagram obtained in the present study. (1) The temperature dependence of $H_{\rm c2}$ in the vicinity of $T_{\rm c}$ is rather gradual. (2) An upward curvature is seen below an inflection point of 2.32 K in the $H_{\rm c2}(T)$ line for $H {\parallel}$ c.

\quad We will below propose that these two features may be explained, at least in a qualitative fashion, by SM's recent theory~\cite{theory}.
As stated in their original paper, they do not intend to consider the mechanism of the enhancement of $T_{\rm c}$ in the eutectic system, but they have constructed a phenomelogical theory. 

\quad The theory includes the following reasonable assumptions: First, 3-K superconductivity occurs at interfaces between Sr$_{2}$RuO$_{4}$ and Ru inclusions. (For simplicity, they treat the interface as a single flat plane, as depicted in the lower panel of Fig. 1.) Second, the superconducting order parameter is represented by a two-component order parameter with a relative phase of $\pi$/2, similar to Sr$_{2}$RuO$_{4}$. 
They used a Ginzburg-Landau (G-L)  free energy for tetragonal symmetry \cite{tetragonal} with the two-component order parameter $\eta = \eta_x + i\eta_y$, corresponding to {\bf{\textit{d}}}(\textbf{\textit{k}}) = {\bf{\textit{z}}}$\Delta_{0}(\eta_{x}k_{x} + i\eta_{y}k_{y})$~\cite{theory}. The G-L free energy also involves a $\delta$-function potential enhancing the $T_{\rm c}$ at the interface between Sr$_{2}$RuO$_{4}$ and Ru.

\quad The above assumptions receive support from existing experimental results such as a weak diamagnetism in the a.c. susceptibility, an imperfect resistance drop mentioned in the introduction and the observation of zero bias conductance peaks~\cite{tunnelling}. 
We have also measured the specific heat for the 3-K phase in the present study. Figure 4 shows the specific heat divided by temperature. A sharp peak is seen at about 1.2 K, which is attributed to the original superconducting transition in Sr$_{2}$RuO$_{4}$. However, a signature of the transition to the 3-K phase is barely observed in the specific heat. This thermodynamically supports the first assumption above. In contrast, the imaginary part of the a. c. susceptibility, displayed in the inset to Fig. 2, shows a broad transition to the 3-K phase well above the sharp 1.5-K original transition. It should be noted that a small hump is seen in the specific heat between 2 and 3 K, which is very close to the transition temperature of the 3-K phase. The attribution of this small hump to the superconducting transiton of the 3-K phase leads to its volume fraction being estimated to be $\sim$1.5{\%}~\cite{3Kvolume}.

\quad Based on the above formulation, SM have considered the upper critical field in fields within the flat interface depicted in the bottom panel of Fig. 1. In both cases of $H {\parallel}$ ab and $H {\parallel}$ c, $H_{\rm c2}$ is proportional to $(1-T/T_{\rm c})^{0.5}$ in the vicinity of $T_{\rm c}$, which is common to surface superconductivity in a field applied parallel to the surface ~\cite{abrikosov}. Examples include superconductivity at twin boundaries ~\cite{abrikosov}.

\quad Fitting the functional form $H_{\rm c2}(T) = A(1-T/T_{\rm c})^n$ to the gradual temperature dependence of $H_{\rm c2}$ in the vicinity of $T_{\rm c}$ yields $n =$ 0.75 and $n =$ 0.72 for $H {\parallel}$ ab and for $H {\parallel}$ c, respectively, where $A$ and $n$ are the adjustable parameters \cite{anotherdef}. These exponents have been obtained from fitting the data between $T_{\rm c}$ and approximately 0.9$T_{\rm c}$. 
As all of these exponents are in contrast to the standard $(1-T/T_{\rm c})$ dependence, we suggest that it supports surface superconductivity in the 3-K phase. 

\quad On the other hand, fitting $(1-T/T_{\rm c})^n$ dependence to the $H$-$T$ phase diagram of the 1.5-K phase based on specific heat measurements \cite {deguchi,deguchi2} yields $n = 0.90$ and $n = 1.0$, for $H {\parallel}$ ab and $H {\parallel}$ c, respectively \cite{fitting}. 
Also a phase diagram from resistive measurements on the 1.5-K phase, albeit the number of data points are rather few, the temperature dependence appears to be linear close to $T_{\rm c}$ \cite{akima}. 
Whilst the exponents of about 0.7 obtained for the 3-K phase somewhat deviate from the predicted value $n = 0.5$, those values are considerably smaller than $n = 1$.

\quad Although we claim that the $H$-$T$ phase diagram obtained probably supports surface superconductivity, a possible criticism is that the exponents obtained being around 0.7 is {\it not} in good enough agreement with the theoretical value of 0.5. 
This discrepancy should originate from the above discussion along the line of SM's theory \cite{theory} being somewhat crude for comparison with experiment. 
Matsumoto and Sigrist \cite{matsumoto} have very recently improved the calculations in SM's paper \cite{theory} and have obtained an exponent of about 0.7 for the temperature range used for our fitting; this exponent is very close to our results. 

\quad The formalism and assumptions Matsumoto and Sigrist have used are in principle based on those of SM's theory. Their important improvement is that Matsumoto and Sigrist use a more realistic wave function of the order parameter in magnetic fields 
than SM's calculations \cite{matsumoto}. 
Whereas SM used an exponetially decaying wave function as in ref. \cite{abrikosov}, Matsumoto and Sigrist have pointed out that this functional form is appropriate only when the field is very low.
In fact, Matsumoto and Sigrist's numerical results indicate that the exponent tends to 0.5 with approaching $T_{\rm c}$ or $H = 0$.
Matsumoto and Sigirst have taken into consideraion the harmonic potential due to the applied magnetic field, leading to a contranction of the wave function. 
They have obtained higher $T_{\rm c}$'s and consequently exponents of around 0.65 and around 0.75 for $H {\parallel}$ ab and for $H {\parallel}$ c, respectively \cite{matsumoto}.

\quad Whilst we intend to discuss the exponent in the vicinity of $H_{\rm c2}$, the range of temperature over which the fit has been applied inevitably has a finite width.
Our exponents quoted above are from a temperature range of approximately 0.9 $T_{\rm c} < T < T_{\rm c}$. 
The exponent from fitting seems not to significantly depend on the temperature range over which the fitting was done, unless the lower temperature limiting the fitting range is too low \cite{inflection} or the number of the data used for the fit are too few \cite{range}.

\quad SM's theory also provides a qualitative explanation for the anomalous behaviour of the $H_{\rm c2}(T)$ curve for $H {\parallel}$ c \cite {theory} ({\it i.e.} upward curvature at low temperatures and high fields). They predict that only one of the two-components of a superconducting order parameter, such as $k_x$ or $k_y$ is stabilised at $T_{\rm c}$ in zero applied field and that the other component with a relative phase of ${\pi}/2$ arises at a slightly lower temperature \cite {similarity}. However, the application of a magnetic field {\it not} perpendicular to the c-axis will induce simultaneously the two components with a relative phase of ${\pi}/2$ at $T_{\rm c}$. (Since the triplet state represented by the order parameter $k_x + i{\varepsilon}k_y (0 < \varepsilon \le 1)$ has an orbital magnetic moment along the c-axis, the state is energetically stabilised by a finite magnetic-field component parallel to the c-axis.) As a consequence of both components being stabilised, the coupling between the two components results in an enhancement of $H_{\rm c2}$. Besides, the coupling becomes stronger at lower temperatures, leading to an upward curvature in the $H_{\rm c2}(T)$ curve for $H {\parallel}$ c. 

\quad In addtion to the machanism described in the last paragraph, Matsumoto and Sigrist suggest that there is another mechanism for the low-temperature high-field enhancement of $H_{\rm c2}$ for $H {\parallel}$ c \cite{matsumoto}.
They have recently raised that 
the region of the enhaced superconductivity (3-K phase) has a finite width in an actual eutectic system although SM adopted the G-L free energy with a $\delta$-function potential enhancing the $T_{\rm c}$ at the interface.
At sufficiently high fields, the spacial extension of the wave function of the order parameter will be cofined within the region where the enhaced superconductivity nucleates, leading to an additional enhancement of $H_{\rm c2}$.

\subsection{Angle Dependence of the Upper Critical Field}
\quad Figure 5(a) shows the angle $\theta$ dependence of the upper critical field $H_{\rm c2}$ at 0.29, 1.32 and 2.45 K. 
($\theta$ is the angle between the ab-plane and the c-axis; $\theta = 0$ corresponds to $H {\parallel}$ ab.) 
Only at 0.29 K of these three temperatures, does $H_{\rm c2}$  show hysteresis close to $H {\parallel}$ ab. For 0.29 K, the hysteresis of $H_{\rm c2}$ persists to 
$|\theta| \approx 10^{\circ}$. (For 60 mK, the hysteresis is observed for $|\theta| \lesssim 20^{\circ}$. The angle range for which the hysteresis can be seen decreases with increasing temperature. As mentioned in subsection A, the hysteresis of $H_{\rm c2}$ disappears at $\sim$1.2 K even for $\theta \approx 0^{\circ}$ ($H {\parallel}$ ab). )
Whilst the lower branch (down sweep of field) for 0.29 K is plotted with solid circles in Fig. 5(a), the upper branch (up sweep of field) is used for the fitting below in this subsection. This is because whether the up sweep or down sweep is used hardly affects the discussion below.
Also shown in Fig. 5(a) are fits of the G-L effective mass model, 
\begin{equation}
H_{{\rm c2}}(\theta) = \frac{H_{{\rm c2} {\parallel} {\rm c}}}{\sqrt{\sin^{2}\theta + {\mathit{\Gamma}}^{-2}\cos^{2}\theta}}.
\end{equation}
Here $\mathit{\Gamma}$ is the square root of the ratio of the effective mass for interplane motion to that for in-plane motion ({\it i.e.} $\mathit{\Gamma} = H_{{\rm c2} {\parallel} {\rm ab}}/H_{{\rm c2} {\parallel} {\rm c}}$) ~\cite {tinkham}. We have taken $H_{{\rm c2} {\parallel} {\rm ab}}$ and $H_{{\rm c2} {\parallel} {\rm c}}$ to be the adjustable parameters for the fitting. The resultant values of $(H_{{\rm c2} {\parallel} {\rm ab}}, H_{{\rm c2} {\parallel} {\rm c}})$ shown in Fig.  5(a) are (3.52 T, 0.92 T), (3.14 T, 0.50 T) and (1.57 T, 0.11 T) for 0.29 K (up sweep), 1.32 K and 2.45 K, respectively. (Those for 0.29 K (down sweep) is (3.33 T, 0.97 T); the curve is not shown.)

\quad Although this model is known to best fit for temperatures close to $T_{\rm c}$, it reproduces as a whole the $\theta$ dependence of $H_{\rm c2}$ for the 1.5-K phase (pure Sr$_2$RuO$_4$) even at 60 mK~\cite{yaguchi}. It should be noted here that a region close to $H {\parallel}$ ab ({\it e.g}. ${\mathit{\Delta}}\theta \le 5^{\circ}$) for the 1.5-K phase is exceptional due to the unusual suppression of the upper critical field; this is probably related to (or caused by) a double superconducting transition~\cite{deguchi,yaguchi}. 
In contrast, Fig. 5(a) exhibits that the model apparently fails to reproduce experimental results of the 3-K phase in a very wide angle range. This discrepancy between the data and the model is particularly evident for low temperature data. 

\quad SM's theory \cite{theory} can be extended to the case of the applied field pointing to arbitrary directions within the Sr$_{2}$RuO$_{4}$-Ru interface plane~\cite{theory,sigrist}. A discussion with a minor simplification yields an analytic functional form identical to the G-L effective mass model (Eq. (1)) \cite{tinkham}. However, this analytic expression is valid for the present system only when the temperature $T$ is close to $T_{\rm c}$ and/or the coupling of the two components is small ({\it i.e.} $H$ is nearly parallel to the ab-plane). In fact, in the framework of the G-L formalism, SM resorted numerical means to investigate the behaviour of $H_{\rm c2}(T)$ for $H {\parallel}$ c at low temperatures.

\quad Consequently, fitting Eq. 1 to data for a certain angle range close to $H {\parallel}$ ab will show a reasonable agreement. In this context, comparison of the data with the effective-mass model will reveal how the discrepancy becomes evident and thus will enable the enhancement of $H_{\rm c2}$ due to the coupling of the two components to be discussed.
As Fig. 5(a) indicates that fitting Eq. 1 to the whole data for $0^{\circ} \le \theta \le 90^{\circ}$ does not yield satisfactory results, we have fitted Eq. 1 to the data for $0^{\circ} \le \theta \le 5^{\circ}$ (at 0.29 and 1.32 K) and $0^{\circ} \le \theta \le 10^{\circ}$ (at 2.45 K), yielding $(H_{{\rm c2} {\parallel} {\rm ab}}, H_{{\rm c2} {\parallel} {\rm c}})$ of (3.62 T, 0.45 T), (3.16 T, 0.41 T) and (1.58 T, 0.11 T) for 0.29 K (up sweep), 1.32 K and 2.45 K, respectively. ((Those for 0.29 K (down sweep) is (3.41 T, 0.46 T))

\quad In Fig. 5(b), the same data as in Fig. 5(a) are plotted as $(H_{\rm c2}(\theta)\cos \theta/H_{\rm c2{\parallel}ab})^2$ vs $(H_{\rm c2}(\theta)\sin \theta/H_{\rm c2{\parallel}c})^2$; the results of the fitting described in the last paragraph are used for $H_{\rm c2{\parallel}ab}$ and $H_{\rm c2{\parallel}c}$ at each temperature.
This plot allows one to see the deviation from the effective mass model more clearly. Since Eq. 1 may be rewritten as 
\begin{equation}
({\frac{H_{\rm c2}(\theta)\cos\theta}{H_{\rm c2{\parallel}ab}}})^2
+ ({\frac{H_{\rm c2}(\theta)\sin\theta}{H_{\rm c2{\parallel}c}}})^2
 = 1, 
\end{equation}
the functional form of Eq. 1 is represented by a straight line connecting (0,1) and (1,0) in this plot.
Figure 5(b) indeed illustrates that Eq. 1 fits well the data at each temperature for a limited angle range close to $H {\parallel}$ ab. The data start to deviate from the functional form of Eq. 1 ({\it i.e.} the dashed straight line in Fig. 5(b)) at about $\theta = 5^{\circ}$ (for 0.29 and 1.32 K) and $\theta = 10^{\circ}$ (for 2.45 K). (Note that the angle $\theta$ at which the deviation becomes evident is irrespective of the choice of values for $H_{\rm c2{\parallel}ab}$ or $H_{\rm c2{\parallel}c}$.)

\quad The deviation is obviously large at low temperatures and large angles (close to $H {\parallel}$ c). In other words, $H_{\rm c2}$ is enhanced at low temperatures and large angles. 
Similarly, ${H_{\rm c2{\parallel}c}}$ from the fitting for the whole data ($0^{\circ} \le \theta \le 90^{\circ}$) is larger than that from the limited range ($0^{\circ} \le \theta \le 5^{\circ}$ or $10^{\circ}$).
This tendency is in very good agreement with SM' s theory~\cite{theory}. They suggest that the coupling between the two components of the order parameter, which enhances $H_{\rm c2}$, becomes stronger with decreasing temperature and increasing magnetic field component parallel to the c-axis.

\quad Before finalising this subsection, we here make a remark on the angle dependende of $H_{\rm c2}$ from another view point. 
The deviation from the G-L effective model Eq. 1 becomes larger with decreasing temperature and $H_{\rm c2}(\theta)$ becomes peaked in the vicinity of $H {\parallel}$ ab at low temperatures. The latter behaviour is somewhat reminiscent of the two-dimensinal thin film model~\cite{tinkham2}, 
\begin{equation}
({\frac{H_{\rm c2}(\theta)\cos\theta}{H_{\rm c2{\parallel}ab}}})^2
+ {\frac{H_{\rm c2}(\theta)|\sin\theta|}{H_{\rm c2{\parallel}c}}}
 = 1. 
\end{equation}
In fact, Eq. 3 shows a peaked feature close to  $H {\parallel}$ ab whilst Eq. 2 does not. 
The thin film model ~\cite{tinkham2} assumes $d \ll \xi_{\rm ab}$ and leads to $H_{{\rm c2} {\parallel} c} =  \frac{\Phi_0}{2\pi\xi_{ab}^2}$ and $H_{{\rm c2} {\parallel} ab} = \frac{\sqrt{3}\Phi_0}{{\pi}d{\xi_{ab}}}$, where $\xi_{\rm ab}$ is the coherence length parallel to the ab-plane, $d$ is the layer spacing and $\Phi_0 = 2.07 \times 10^{-15}~{\rm T}/{\rm m}^2$ is the fluxoid. Nevertheless, the use of these formulae for the 3-K phase at 60 mK results in $\xi_{\rm ab} =$ 16.2 nm and $d =$ 1.90 nm, which does not satisfy a prerequisite of the model, $d \ll \xi_{\rm ab}$. Also, $d$ = 1.90 nm is substantially larger than the layer spacing of Sr$_2$RuO$_4$, 0.637 nm. These facts suggest the application of the thin film model to the 3-K phase is inappropriate.

\section{Conclusion}
\quad In summary, we have investigated the field-temperature phase diagram of the 3-K phase of Sr$_2$RuO$_4$ in detail using resistivity measurements. We have found a rather gradual temperature dependence of the upper critical field $H_{\rm c2}$ close to $T_{\rm c}$ and an enhancement of $H_{\rm c2}$ for $H {\parallel}$ c at low temperatures. We have also investigated the dependence of $H_{\rm c2}$ on the angle between the field and the ab-plane at several temperatures. Fitting of the G-L effective-mass model apparently fails to reproduce the angle dependence. 
All of these experimental results, with the help of the theory of SM may be interpreted in a consistent manner with other existing experimental facts. Taken together with the phenomelogical theory by SM, these observations support that the 3-K phase is surface spin-triplet superconductivity with a two-component order parameter occurring at Sr$_{2}$RuO$_{4}$-Ru interfaces. This, although indirectly, supports the basic form of {\bf{\textit{d}}}(\textbf{\textit{k}}) = {\bf{\textit{z}}}$\Delta_{0}(k_x + ik_y)$ for bulk Sr$_2$RuO$_4$ as well.

\begin{acknowledgements}
\quad We thank M. Sigrist, M. Matsumoto, H. Monien and K. Deguchi for invaluable discussions and enlightening suggestions. 
M. Matsumoto is also thanked for access to unpublished results prior to publication.
We are grateful to Y. Mori and T. Ando for their contribution to the present work at early stages. This work was in part supported by the Grant-in-Aid for Scientific Research (S) from the Japan Society for Promotion of Science and by the Grant-in-Aid for Scientific Research on Priority Area "Novel Quantum Phenomena in Transition Metal Oxides" from the Ministry of Education, Culture, Sports, Science and Technology.
\end{acknowledgements}

\begin{figure} 
\caption{Top: Optical microscopy picture of a polished surface parallel to the RuO$_{2}$ plane (bright region: Ru, dark region: Sr$_2$RuO$_4$).
Bottom: Schematic of the interface between Sr$_2$RuO$_4$ and a Ru-inclusion modelled in Sigrist and Monien's theory. The interface within the Sr$_2$RuO$_4$ part has a thin layer where a p-wave state nucleates at an enhanced transition temperature of $\sim$ 3 K; its wavefunction has lobes and nodes parallel and perpendicular to the interface, respectively.}
\end{figure}

\begin{figure} 
\caption{Typical traces of the resistance of the 3-K phase (solid lines) and their derivatives with respect to magnetic field or temperature (dashed lines). These illustrate that the transition points (inflection points) from field sweep and temperature sweep show a good agreement.}
\end{figure}

\begin{figure} 
\caption{Field-temperature phase diagram of the 3-K phase. The transition points have been determined as the inflection point associated with the superconducting transition to the 3-K phase. The $H_{\rm c2}$ curve for $H {\parallel}$ c shows an upward curvature with an inflection of 2.32 K, as indicated by an arrow. The dashed curves represent fits of $(1-T/T_{\rm c})^n$ dependence to data close to $T_{\rm c}$. ($n = 0.75$ and 0.72 for $H {\parallel}$ ab and $H {\parallel}$c, respectively)}
\end{figure}

\begin{figure}
\caption{Specific heat devided by temperature $C_p / T$ of the  eutectic system Sr$_{2}$RuO$_{4}$-Ru. Whilst a clear peak associated with the 1.5-K superconducting transition is seen, a signature of 3-K supeconductivity is barely evidenced. Inset: imaginary part of the a. c. susceptibility. A broad feature associated with the superconducting transiton to the 3-K phase is seen.}
\end{figure}

\begin{figure} 
\caption{(a) Angle $\theta$ dependence of the upper critical field $H_{\rm c2}$ at 0.29, 1.32 and 2.45 K. The dashed curves represent fits of the G-L effective mass model for $0^{\circ} \le \theta \le 90^{\circ}$. 
For 0.29 K, there are two branches reflecting hysteresis of $H_{\rm c2}$; the open (solid) circles correspond to up (down) sweep of field.
(b) The same data (but without down-sweep branch at 0.29 K) and fits of the G-L effective mass model for $0^{\circ} \le \theta \le 5^{\circ}$ (for 0.29 and 1.32 K) and $0^{\circ} \le \theta \le 10^{\circ}$ (for 2.45 K). The data are plotted as $(H_{\rm c2}\cos \theta/H_{\rm c2{\parallel}ab})^2$ vs $(H_{\rm c2}\sin \theta/H_{\rm c2{\parallel}c})^2$, so that all of the fits are represented by the dashed straight line.}
\end{figure}


\begin {references}
\bibitem{3Kvolume} 
The estimation of the 3-K phase volume fraction is based on the ratio of 
$\frac{{\mathit{\Delta}C_{p}}}{{\gamma}_{\rm N}T_{\rm c}}$ of the 3-K phase to that of the 1.5-K phase, with the value of ${\gamma}_{\rm N}$ taken to be common.
\bibitem{anotherdef}
It seems that the exponents do not strongly depend on definition of $T_{\rm c}$ (or $H_{\rm c2}$). We have also used another definition for $T_{\rm c}$ (or $H_{\rm c2}$): the onset of the superconducting transition. The exponents $n$ from this definition are $n =$ 0.67 and $n =$ 0.70 for $H {\parallel}$ ab and for $H {\parallel}$ c, respectively.
\bibitem{fitting}
The fittings have been done for a temperature range of approximately $0.9 T_{\rm c} < T < T_{\rm c}$, similar to the case of the 3-K phase in the present work.
\bibitem{inflection}
This is particularly obvious for $H {\parallel}$ c as there is an inflection point at $T$ = 2.32 K.
\bibitem{range}
When the lower temperature limiting the fitting range is between 2.4 K (0.85$T_{\rm c}$) and 2.6 K (0.93$T_{\rm c}$), the resultant exponent hardly changes. The exponent for $H {\parallel}$ ab stays 0.75; that for $H {\parallel}$ c lies between 0.72 and 0.77. The calculations by Matsumoto and Sigrist suggest that the exponent is between 0.61 and 0.63 for $H {\parallel}$ ab, and is between 0.75 and 0.77 for $H {\parallel}$ c.
\end{references}
\end{document}